\documentclass{elsart}

\usepackage{epsfig}

\usepackage{amssymb}

\begin{document}

\begin{frontmatter}



\title{Multi-component power spectra estimation method for multi-detector observations of the Cosmic Microwave Background}



\author{Guillaume Patanchon}

\address{PCC --- Coll{\`e}ge de France, 11, place Marcelin Berthelot, F-75231 Paris, France}

\begin{abstract}

We present a new method for multi-component power spectra estimation in
multi-frequency observations of the CMB. Our method is based on matching
the cross and auto power spectra of observation maps to their expected values.
All the component
power spectra are estimated, as well as their mixing matrix. Noise power
spectra are also estimated. The method has been applied to full-sky Planck
simulations containing five components and white noise. The beam smoothing
effect is taken into account.

\end{abstract}

\end{frontmatter}

\section{Introduction}
\label{}

Precise measurements of Cosmic Microwave Background (CMB) angular power
spectrum, is one of the main objectives of modern Cosmology. Several
experiments, such as ACBAR \cite{ACBAR}, Archeops\cite{Archeops}, Boomerang
\cite{Boomerang}, CBI \cite{CBI}, DASI \cite{DASI}, MAXIMA \cite{MAXIMA}, VSA
\cite{VSA} and more recently WMAP \cite{WMAP} already give precise results over
a large range of angular
scales. An unbiased estimate of the CMB angular power spectrum requires
important efforts in data processing because of the presence of several sources
of contamination in the observations. Foregrounds, in particular
(Galactic dust, synchrotron, extragalactic point sources and Sunyaev-Zel'dovich
effects), will be the main contaminants for CMB measurements, such as that of
the future Planck mission. Fortunately, the avaibility of several detectors
operating at several frequencies, ranging from 30 GHz to 857 GHz for Planck,
permit the separation of most astrophysical components due to their different
emission laws.\\ Starting from the different observation maps, current methods
for CMB power spectrum estimation consist first in making the cleanest CMB
anisotropy map possible, using component separation techniques, like Wiener
filtering \cite{Wiener,wiener96tegmark,allskywienerPrunet} or MEM
\cite{MEM,allskyMEM}. Spectral estimation is then performed on the individual
component maps. This approach is not fully satisfactory for two reasons: First,
component separation requires prior knowledge of the electromagnetic spectra of
the components, which are not necessarily very well known. Second, after
component separation, in the subsequent spectral estimation step, it is
necessary to remove the power spectrum of the residual noise in the extracted
CMB map. A small error in the noise evaluation introduces a bias in CMB power
spectrum estimation.\\ A new approach has been considered in paper
\cite{MDMC}. This approach is based on the analysis of cross- and auto-power
spectra of different observation maps by different detectors. Considering the
simple case where only CMB anisotropies are present with noise, the cross-power
spectra give a direct measurement of the CMB power spectrum, assuming that the
noise is uncorrelated between detectors. In this limited context, the technique
has been used by the WMAP team. Our approach deals simultanously with
multi-component data. By jointly analysing the different observation maps, we
estimate the power spectra of all the components, without making any priors on
their emission laws. Thus, the contribution of each component to each detector,
given by the mixing matrix $A$ (e.g. (1)), is estimated directly from the data.
The noise power spectra are also estimated. The method is based on likelihood
maximization in the Whittle approximation assuming that the observations are a
linear mixing of independant components and independant noise. It can be viewed
as a blind multi-detector multi-component (MDMC) spectral matching, in which
the spectral diversity of the various components is used. The method has been
implemented on harmonic coefficients of all sky maps. We account for the finite
spatial resolution of the detectors. In this paper, we describe the basics of
the spectral matching method and we present its application on full-sky
simulated Planck observations.

\section{Model for the observed sky}
\label{}

The key assumptions are that the sky emission at a given frequency is a linear
superposition of astrophysical components, and that their emission laws do not
depend on sky position. The signal measured by a detector is the sky emission
convolved by a beam shape (depending on the detector), plus an additive noise.
A spherical harmonic expansion over a full sky map, observed by detector $d$,
gives us:
\begin{equation}
 {x_d}(\ell,m) = b_d(\ell) \sum_{c=1}^{N_c} A_{dc}~s_c(\ell,m) + n_d(\ell,m)
\label{eq:obsmodel}
\end{equation}

where $s_{c}$ is the emission template for source $c$, $n_d$ represents the
noise and $A$ is the mixing matrix. Each element of the mixing matrix results
from the integration of the emission law of one component over one detector
frequency band. The coefficient $b_d(\ell)$ refers to a Legendre Polynomial
expansion of the beam, depending only on $\vec{r}.\vec{r'}$ (we assume a
symmetric beam). Let us consider the diagonal matrix $B(\ell)$ such that the
diagonal element $B_{dd}(\ell) = b_d(\ell)$. It is useful to define the
``deconvolved'' observation coefficients ${x'}(\ell,m)=B(\ell)^{-1} x(\ell,m)$,
that we write in matrix form:
\begin{equation}
  {x'}(\ell,m) = A s(\ell,m) + B(\ell)^{-1}n(\ell,m)
\label{eq:ourmodel}
\end{equation}

{\bf{Observation power spectra}}

Our method is based on the analysis of cross- and auto-power spectra of the
``deconvolved'' observation maps, whose ensemble averages are expressed as~:
\begin{equation}
  R_{x'}(\ell) = {1 \over 2\ell+1}\sum_{m=-\ell}^{\ell} \langle x'(\ell,m) x'(\ell,m){^\dagger} \rangle.
\end{equation}
where $\cdot^\dagger$ denotes transpose-conjugation. We define $C(\ell)$ and
$N(\ell)$ as the component and the noise power spectra respectively, we assume
statistical independence between components and also between the noise of
different detectors, so $C(\ell)$ and $N(\ell)$ are diagonal matrices. The
expected observation power spectra become, using (\ref{eq:ourmodel}):
\begin{equation}
  R_{x'}(\ell) = A C(\ell) A^t + M(\ell)
\label{eq:sptheo}
\end{equation}
where $M(\ell) = B(\ell)^{-2}N(\ell)$ is also a diagonal matrix.\\
Observation power spectra are estimated in the data by:
\begin{equation}
  \widetilde R_{x'}(\ell) = {1 \over 2\ell+1} \sum_{m=-\ell}^{\ell} {x'}(\ell,m){x'}(\ell,m)^\dagger
\label{eq:spexp}
\end{equation}
Expected and measured power spectra can be averaged over bins $q$ such that,
for example, $\ell_{\rm min}(q)<\ell<\ell_{\rm max}(q)$. Their structures, as
in equations (\ref{eq:sptheo}) and (\ref{eq:spexp}) are preserved:
\begin{eqnarray}
  R_{x'}(q) & = & A C(q) A^t + M(q) \label{eq:bintheo}\\
  \widetilde R_{x'}(q) & = & {1 \over n_q} \sum_{\ell=\ell_{\rm min}(q)}^{\ell_{\rm max}(q)}\sum_{m=-\ell}^{\ell} {x'}(\ell,m){x'}(\ell,m)^\dagger \label{eq:binexp}
\end{eqnarray}
where $n_q$ is the number of modes in bin $q$. According to equation
(\ref{eq:bintheo}), the non-diagonal part of $R_{x'}(q)$ contains only
contributions from components (by supposition of noise independance). This
shows that the cross power spectra gives a direct estimate of the mixing of
component power spectra; this mixing can be inverted, even without priors on
$A$, assuming independance of components and some spectral diversity.

\section{Multi-component spectral matching}
\label{}

Our method is based on minimizing the mismatch between the empirical power
spectra of the observations (\ref{eq:binexp}) and their expected values
(\ref{eq:bintheo}), in order to measure simultenaously the averaged component
power spectra $C(q)$ and the mixing matrix $A$ (see discussion in paper
\cite{Ppatanch} on a blind vs a semi-blind approach). As we wish to avoid
using any priors, the averaged noise power spectra $M(q)$ are also estimated.
These parameters are referred to as $\theta=\{A,C(q),M(q)\}$. The mismatch is
quantified by the average divergence measure between the two matrices:
\begin{equation}
   \Phi(\theta) = \sum_{q=1}^Q n_q D(\widetilde R_{x'}(q),R_{x'}(q))
\label{equation:phi}
\end{equation}
Assuming that the spherical harmonic coefficients of the components are
random realizations of a Gaussian field of variance $R_{x'}(q)$, and that they
are all uncorrelated, the log-likelihood (up to an irrelevant factor) takes the
same form as in equation (\ref{equation:phi}) (in the frame of the Whittle
approximation), in this case the divergence is given by
$D(R_1,R_2) = {\rm tr} (R_1R_2^{-1}) - {\rm log~det} (R_1R_2^{-1})-m_d$
\cite{MDMC}. The estimate $\hat\theta$ is such that $\Phi(\theta)$ is minimum. The
connection with the log-likelihood garanties asymptotically minimum variance of
the estimates and the absence of bias (this last property is preserved even if
the components and noise are non-stationnary and non-Gaussian).\\
In practice, we optimize the criteria $\Phi(\theta)$ in a first step using an
EM (Expectation-Maximization) algorithm and the convergence is ended in a
second step using a Quasi-Newton method (see papers \cite{MDMC,Pcardoso} for a
description of the algorithm).


\section{Application to Planck observation simulations}
\label{}

\begin{figure}[tb]
  \centering
\centerline{\epsfig{figure=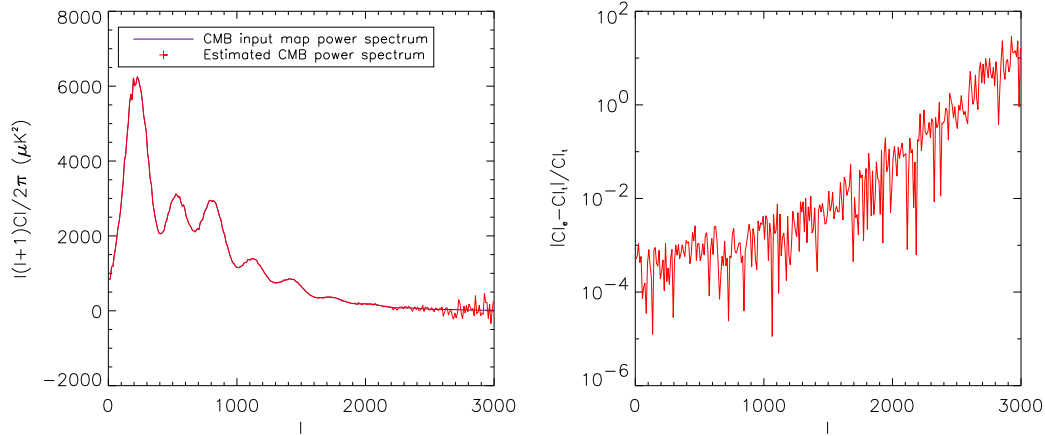,width=14cm}}
\caption{Left: Estimated CMB power spectrum and CMB power spectrum of the input
map. Right: Relative errors made on the CMB power spectrum estimation.}
\label{fig:res}
\end{figure}

We now turn to the application of MDMC spectral matching method on simulated
all-sky Planck observations provided by the Planck consortium. The maps are
generated for the ten frequency channels of the Planck instruments. Five components~:
CMB, thermal dust, the two SZ effects and synchrotron, and white noise at the
nominal instrument level are mixed according to equation (\ref{eq:obsmodel}).
The beam sizes are also taken at their nominal values. See paper
\cite{allskyMEM} for more details on the simulations.\\
The cross- and auto-power spectra of simulated Planck observations are computed
following equation (\ref{eq:binexp}), up to the multipole $\ell = 3000$ and
using bins of width $\Delta l = 10$. We try to estimate four components.
Indeed, the kinetic SZ can not be separated from the CMB anisotropies with our
method since these two components have proportional emission laws (they form
one component). We estimate all component power spectra and all mixing matrix
elements except three, corresponding to three of the four elements at 857GHz,
which are fixed at zero (we expect only thermal dust at this frequency). This
operation does not affect CMB power spectrum estimation results, but allows us
to break degeneracies between galactic components having almost proportional
power spectra. We assume as prior information that the noise is white ($N$
independant of $\ell$). The total number of parameters is $10 \times 4 -3 +4
\times 300+10 = 1247$, compared to $300 \times 10 \times (10 + 1)/2 = 16500$
data elements.\\
The mixing parameters relative to the four components are estimated with
good accuracy (see paper \cite{Ppatanch}). In particular, we are able to put
strong constrains on emission laws of Galactic components. Figure \ref{fig:res}
shows the estimated CMB power spectrum and the estimated relative errors given
by ($|\tilde C(q) - C(q)|/C(q)$). It appears that the method allows us to
accurately estimate the CMB power spectrum up to $\ell \simeq 2000$. No bias in
the estimation is seen over the entire $\ell$ multipole range.

\section{Discussion}

{\bf{Single component}}

The sensitivity of current CMB experiments lead us to expect that at very
high galactic latitude at frequencies around 150 GHz, other astrophysical
components are negligible as compared to CMB and noise (this will not be the
case for Planck). Even in this simple case, our method, applied on small sky
regions observed by multi-detector experiments, gives comparable or even better
CMB power spectrum estimates than the ``classical'' techniques. The advantage
of our method is that an independant noise power spectra estimation is not
required since we explicitly consider cross power spectra of observation maps.

{\bf{Component map separation}}

Our spectral matching method yields all the parameters needed to implement a
Wiener-based or MEM component separation on the maps. Note that we proceed in
making a spectral estimation followed by a component separation, instead of the
inverse, as curently done in classical approaches. We have applied Wiener
filtering on simulated Planck observations using the estimated parameters.
Component maps are accurately estimated, in particular output CMB map does not
show residual contaminations in galactic plane regions. 

\section{Conclusion}
\label{}

We have presented our blind spectral matching method for all-sky multi-detector
CMB observations. The main objective is to measure the power spectra of all the
components, including the CMB, and their contribution levels at each
observation frequency. The method exploits in particular the cross-power
spectra of observation maps, giving an unbiased estimate of component power
spectra. The method has been applied on full-sky Planck simulations containing
five components and white noise. The power spectrum of the CMB is accurately
estimated up to $l \simeq 2000$ in bins of size $\Delta l = 10$.

{\bf{Acknowledgments.}}
The author would like to thank the Planck collaboration and in particular
M.A.J. Ashdown, V. Stolyarov and R. Kneissl for the full sky simulated maps,
and also J. Bartlett for a critical reading of the manuscript.
The HEALPix package (see http://www.eso.org/ science/healpix/) was used for the
spherical harmonics decomposition of the input maps.




\end{document}